\documentstyle[graphics,epsf]{llncs}
\newcommand{\be}{\begin{equation}}
\newcommand{\ee}{\end{equation}}
\newcommand{\bea}{\begin{eqnarray}}
\newcommand{\eea}{\end{eqnarray}}
\newcommand{\nn}{\nonumber}

\newcommand{\ket}[1]{{|#1\rangle}}

\begin{document}

\title{Trapped Ion Quantum Computer Research at Los Alamos}

\author{D. F. V. James, M. S. Gulley, M. H. Holzscheiter, 
R. J. Hughes, P. G. Kwiat, S. K. Lamoreaux, C. G. Peterson,
V. D. Sandberg, M. M. Schauer,\\ C. M. Simmons, D. Tupa,
P. Z. Wang, A. G. White.}

\institute{ Los Alamos National Laboratory, Los Alamos, NM 87545, USA }

\maketitle
\begin{abstract}
We briefly review the development and theory of an experiment 
to investigate quantum computation with
trapped calcium ions. The ion trap, laser and ion requirements
are determined, and the parameters required for simple quantum logic 
operations are described. \\(LAUR 98-314)
\end{abstract}

\section{Introduction}
In the last 15 years various authors have considered the generalization
of information theory concepts to allow the representation of information by
quantum systems. The introduction into computation of {\em quantum mechanical} 
concepts, in particular the superposition principle, 
opened up the possibility of new capabilities, such as quantum
cryptography \cite{HughesCrypto}, that have no classical counterparts. 
One of the most
interesting of these new ideas is quantum computation, first proposed by
Benioff \cite{Benioff}.  Feynman  \cite{Feynman} suggested
that quantum computation might be
more powerful than classical computation, a notion which gained further
credence through the work of Deutsch  \cite{Deutsch}. 
However, until quite recently
quantum computation was an essentially academic endeavor because there
were no quantum algorithms that exploited this power to 
solve useful computational problems, and because no realistic
technology capable of performing quantum computations had been envisioned. 
This changed in 1994 when Shor discovered quantum
algorithms for efficient solution of integer
factorization and the discrete logarithm problem 
\cite{Shor,EkertJozsa}, two problems that are at the heart of
the security of much of modern public key cryptography
\cite{Hughes}. Later that same year 
Cirac and Zoller proposed that quantum computational hardware could be
realized using known techniques in the laser manipulation of trapped 
ions \cite{CZ}.
Since then interest in quantum computation has grown dramatically, 
and remarkable progress has been made: a single quantum logic gate
has been demonstrated with trapped ions \cite{NISTgate};
quantum error correction schemes
have been invented \cite{MannyRay,Preskill};
several alternative technological proposals have been made 
\cite{Kimble,Havel2,Ziolo,Privman,Bocko,DiVincenzo2}
and quantum algorithms for solving new problems have
been discovered \cite{Grover,Terhal,Boneh,Kitaev}. 
In this paper we will review  our development of an experiment 
to investigate the potential of quantum computation 
using trapped calcium ions \cite{latiqce}.

The three essential requirements for quantum computational hardware are: 
(1) the ability to isolate a set of two-level quantum systems from 
the environment for long enough to maintain 
coherence throughout the computation, while at the same time
being able to interact with the systems strongly enough to
manipulate them into an arbitrary quantum state; 
(2) a mechanism for performing quantum logic operations: in
other words a ``quantum bus channel'' connecting the various
two-level systems in a quantum mechanical manner; and
(3) a method for reading out the quantum state of the system
at the end of the calculation.

\begin{figure}[!ht]
\begin{center}
\epsfxsize=8cm  
\epsfbox{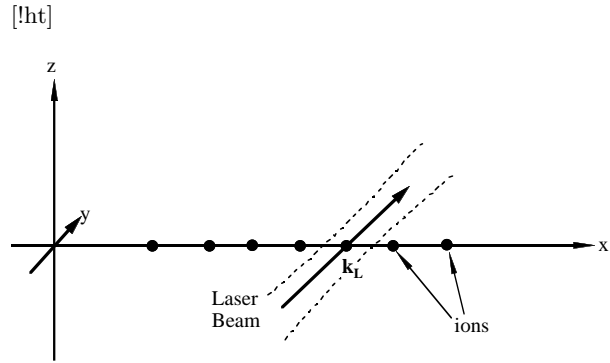}
\end{center}
\caption{A schematic illustration of an idealized laser-ion interaction 
system; ${\bf k}_{L}$ is the wavevector of the single
addressing laser.}
\label{figone}
\end{figure}

All three of these requirements are in principle met by
the cold trapped ion quantum computer.
In this scheme each qubit 
consists of two internal levels of an ion trapped in a linear 
configuration.  In order to perform the
required logic gates, a third atomic state
known as the auxiliary level is required. The 
quantum bus channel is realized using the phonon modes of the ions'
collective oscillations. These quantum systems
may be manipulated using precisely controlled
laser pulses.  Two distinct types of laser pulse
are required: ``V'' type pulses, which only interact
with the internal states of individual ions, and
``U'' type pulses which interact with both the
internal states and the external vibrational 
degrees of freedom of the ions.  These interactions
can be realized using  Rabi flipping
induced by either a single laser or Raman (two laser)
scheme (Fig.2). Readout is performed by
using quantum jumps.  This scheme was originally proposed 
by Cirac and Zoller in 1994 \cite{CZ}, and was used to demonstrate a 
CNOT gate shortly afterwards \cite{NISTgate}.

\begin{figure}[!ht]
\begin{center}
\epsfxsize=7.5cm  
\epsfbox{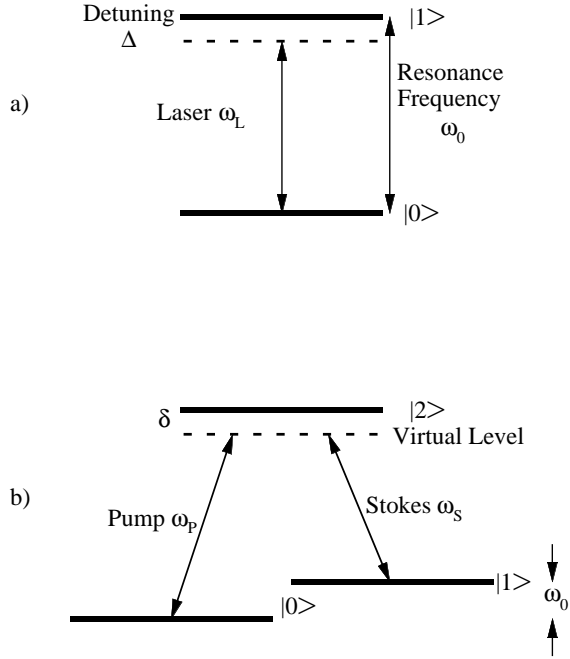}
\end{center}
\caption{A schematic illustration of (a) single laser and
(b) Raman qubit addressing and control techniques.}
\label{figtwo}
\end{figure}

As we can only give the briefest of description
of the principles of quantum computation using cold
trapped ions, the reader is recommended to peruse the more
detailed descriptions which can be found elsewhere
\cite{Steane,dfvj,NISTrev,latiqce}.  In this paper
we intend to focus on the experimental issues involved
in building a trapped ion quantum computer.
\section {Choice of Ion}
There are three requirements which the species of ion
chosen for the qubits of an ion trap quantum computer
must satisfy:

1. If we use the single laser scheme, the ions must have a level 
that is sufficiently long-lived to allow
some computation to take place; this level can also be used for
sideband cooling.

2. the ions must have a suitable dipole-allowed transition for Doppler
cooling, quantum jump readout and for Raman transitions (if we chose
to use two sub-levels of the ground state to form the qubit);

3.  These transitions must be at wavelengths compatible with current
laser technology.

\begin{figure}[!ht]
\begin{center}
\epsfxsize=8.4cm  
\epsfbox{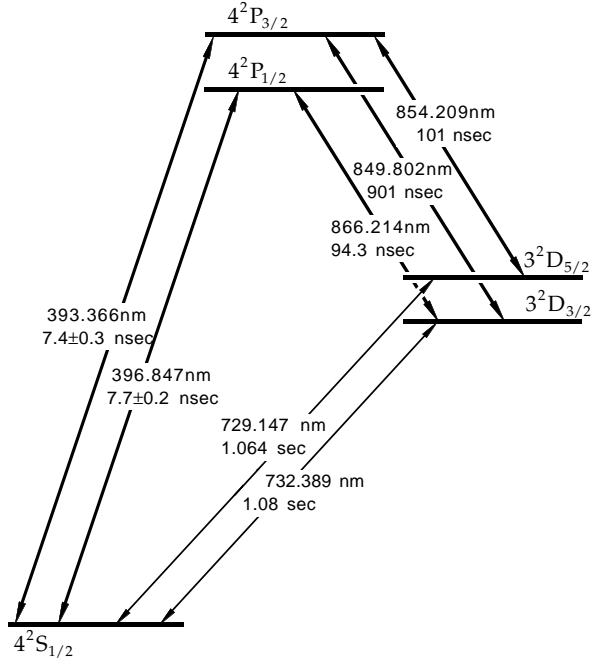}
\end{center}
\caption{The lowest energy levels of $^{40}Ca^{+}$ ions, 
with transition wavelengths and lifetimes listed.}
\label{figthree}
\end{figure}

Various ions used in atomic frequency standards work
satisfy the first requirement.
Of these ions, $Ca^{+}$ offers the advantages of transitions
that can be accessed with titanium-sapphire or diode lasers and a reasonably
long-lived metastable state. 
The relevant energy levels of the $A=40$ isotope are shown in fig.3. 

The dipole-allowed transition from the $4\,^{2}S_{1/2}$ ground state to the 
$4\,^{2}P_{1/2}$
level with a wavelength of 397 nm can be used for Doppler cooling and
quantum jump readout; The 732 nm electric quadrupole transition from the 
$4\,^{2}S_{1/2}$ ground state to the $3\,^{2}D_{3/2}$ metastable level 
(lifetime $\approx 1.08 sec.$) is
suitable for sideband cooling. 
In the single laser computation scheme,
the qubits and auxiliary level can be chosen
as the electronic states
\bea
\ket{0} &=&\ket{4\,^{2}S_{1/2},\,M_{j}=1/2}, \nn \\ 
\ket{1} &=&\ket{3\,^{2}D_{5/2},\,M_{j}=3/2}, \nn \\ 
\ket{aux}&=&\ket{ 3\,^{2}D_{5/2},\,M_{j}=-1/2} . \nn
\eea

This ion can also be used for Raman type qubits, with the two
Zeeman sublevels of the $4\,^{2}S_{1/2}$ ground state forming the
two qubit states $\ket{0}$ and $\ket{1}$, 
with one of the sublevels of the $4\,^{2}P_{1/2}$ level
being the upper level $\ket{2}$.  A magnetic field of
200 Gauss should be sufficient to split these two levels
so that they can be resolved by the lasers.  The pump and Stokes beams 
would be formed by splitting a 397nm laser into two, and
shifting the frequency of one with respect to the other
by means of an acousto-optic or electro-optic modulator.  This arrangement
has a great advantage in that any fluctuations in the
phase of the original 397nm laser will be passed on to
both the pump and Stokes beams, and will therefore be
canceled out, because the dynamics is only sensitive
to the difference between the pump and Stokes phases.
One problem in realizing the Raman scheme in $Ca^{+}$
is the absence of a third level in the ground state that
can act as the auxiliary state $\ket{aux}$ required for
execution of quantum gates.  This difficulty could be removed
by using the alternative scheme for quantum logic recently
proposed by Monroe {\it et al.} \cite{NISTsimp}; alternatively, one could 
use an isotope of $Ca^{+}$ which has non-zero nuclear spin, thereby
giving several more sublevels in the ground state due to the hyperfine
interaction; other possibilities that have been suggested
for an auxiliary state with $^{40}Ca^{+}$ in the Raman scheme
are to use a state of a phonon mode other than the CM mode 
\cite{Steaneaux}
or one of the sublevels of the $3\,^{2}D$ doublet \cite{Blattaux}.
\section {The Radio Frequency Ion Trap}
Radio-frequency (RF) quadrupole traps, also named ``Paul traps'' after their
inventor, have been used for many years to confine
electrically charged particles \cite{Paul} (for an introduction
to the theory of ion traps, see refs. \cite{NISTtrap,Ghosh}).
The classic design of such a Paul trap
has a ring electrode with endcap electrodes above and below, with
the ions confined to the enclosed volume. A single ion can be located
precisely at the center of the trap where the amplitude of the RF field is
zero. But when several ions are placed into this trapping field, their
Coulomb repulsion forces them apart and into regions where they are
subjected to heating by the RF field. For this reason in our experiment
ions are confined in a linear RF quadrupole trap \cite{latiqce}.
Radial confinement is achieved by a quadrupole RF
field provided by four 1 mm diameter rods in a rectangular
arrangement.  Axial confinement is provided by DC voltages applied
to conical endcaps at either end of the RF structure; the endcap
separation is 10 mm.
The design of the trap used in these experiments is shown in
diagrammatically in Fig.4.  

\begin{figure}[!ht]
\begin{center}
\epsfxsize=8cm  
\epsfbox{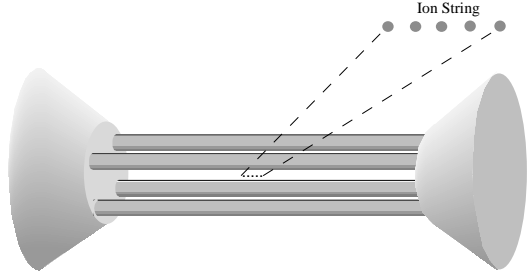}
\end{center}
\caption{Side view diagram of the linear RF trap used to confine
$Ca^{+}$ ions in these experiments.  The endcap separation is 10 mm
and the gap between the RF rods is 1.7 mm.}
\label{figfour}
\end{figure}
 
The main concerns for the design are to provide sufficient
radial confinement to assure that the ions form a string on the
trap axis after Doppler cooling; 
to minimize the coupling between the radial and axial
degrees of freedom by producing radial oscillation frequencies
significantly greater than the axial oscillation frequencies;
to produce high enough axial frequencies to allow the use of
sideband cooling\cite{NISTcool89}; and to provide sufficient spatial
separation
to allow individual ions to be addressed with laser beams.
\section {Laser Systems}
The relevant optical transitions for $Ca^{+}$ ions are shown in 
Fig.5.  There are 
four different optical processes employed in the quantum computer; each places 
specific demands on the laser system.

\begin{figure}[!ht]
\begin{center}
\epsfxsize=8cm  
\epsfbox{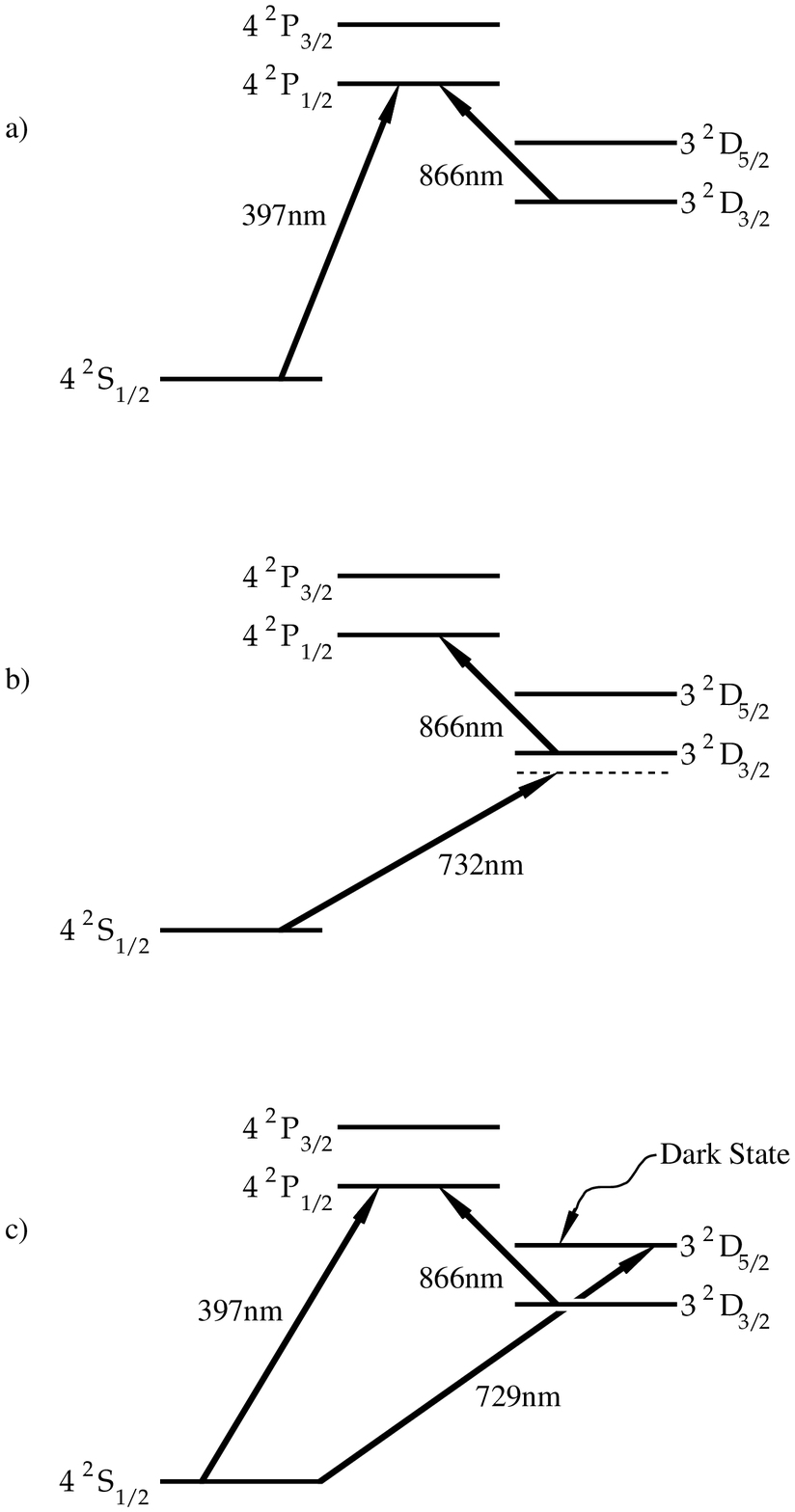}
\end{center}
\caption{Different transitions between the levels
of $Ca^{+}$ ions required for (a) Doppler cooling,
(b) Resolved sideband cooling and (c) quantum logic operations
and readout.  The single laser addressing technique
has been assumed.}
\label{figfive}
\end{figure}

The first stage is to cool a small number of ions to their Doppler limit in the ion 
trap, as shown in Fig.5a.  This requires a beam at 397 nm, the 
$4\,^{2}S_{1/2} - 4\,^{2}P_{1/2}$
resonant transition.  Tuning the laser to the red of the transition causes the 
ions to be slowed by the optical molasses technique \cite{StenholmCool}.  
In this procedure, a 
laser beam with a frequency slightly less than that of the resonant transition of an 
ion is used to reduce its kinetic energy.  Owing to the Doppler shift of the photon 
frequency, ions preferentially absorb photons that oppose their 
motion, whereas they re-emit photons in all directions, 
resulting in a net reduction in momentum along the direction of the
laser beam.  Having carefully selected the trap 
parameters, many cycles of absorption and re-emission will bring the system to 
the Lamb-Dicke regime, leaving the ions in a string-of-pearls geometry.
We have recently found ion crystals of up to five $Ca^{+}$ ions.

In order to Doppler cool the ions, the demands on the power and linewidth of the 
397 nm laser are modest.  The saturation intensity of $Ca^{+}$ ions is 
$\sim 10\; {\rm mW/cm}^{2}$, and the laser linewidth must be less than $\sim 
10\; {\rm MHz}$. 
 An optogalvonic 
signal obtained with a hollow cathode lamp suffices to set the frequency.  We 
use a Titanium:Sapphire (Ti:Sapphire) laser (Coherent CR 899-21) with an internal 
frequency doubling crystal to produce the 397 nm light.

During the Doppler cooling, the ions may decay from the $4\,^{2}P_{1/2}$ state 
to the $3\,^{2}D_{3/2}$
state, whose lifetime is $\sim 1 {\rm sec}.$  
To empty this metastable state, we use a second 
Ti:Sapphire laser at 866 nm.  

Once the string of ions is Doppler cooled to the Lamb-Dicke regime, the second 
stage of optical cooling, sideband cooling, will be used to reduce the collective 
motion of the string of ions to its lowest vibrational level 
\cite{WinelandItano}, illustrated in Fig.5b.  
In this regime, a narrow optical transition, such as the 732 nm 
$4\,^{2}S_{1/2} - 3\,^{2}D_{3/2}$ dipole forbidden transition,
develops sidebands above and below the 
central frequency by the vibrational frequencies of the ions.  The sidebands 
closest to the unperturbed frequency correspond to the CM vibrational motion.  
If $\omega_{0}$ is the optical transition frequency and $\omega_x$ 
the frequency of the CM 
vibrational motion,  the phonon number is increased by one, unchanged, or 
decreased by one if an ion absorbs a photon of frequency 
$\omega_{0}+\omega_x$, 
$\omega_{0}$ or $\omega_{0}-\omega_x$, 
respectively.  Thus, sideband cooling is accomplished by optically cooling the 
string of ions with a laser tuned to $\omega_{0}-\omega_x$.  

The need to resolve the sidebands of the transition implies a much more stringent 
requirement for the laser linewidth; it must be well below the CM mode vibrational 
frequency of $\sim  (2\pi)\times 200\; {\rm kHz}$.  The laser power 
must also be greater in order to pump 
the forbidden transition.  We plan to use a Ti:Sapphire 
laser locked to a reference cavity to 
meet the required linewidth and power.  At first glance it would seem that, with 
a metastable level with a lifetime of 1s, no more than 1 phonon per second could 
be removed from a trapped ion.  A second laser at 866 nm is used to couple the 
$4\,^{2}P_{1/2}$ state to the $3\,^{2}D_{3/2}$ state to reduce the effective 
lifetime of the D state and 
allow faster cooling times.  The transitions required for realization 
of quantum logic gates and for readout, discussed in detail in
sections 5.2 and 5.3, are shown in Fig.5c.  These
can be performed with the same lasers used in the Doppler and sideband
cooling procedures.

There are two other considerations concerning the laser systems for 
quantum computation which should be mentioned.  To reduce the total
complexity of the completed system, we are 
developing diode lasers and a frequency doubling cavity to handle the Doppler 
cooling and quantum jump read out.  Also complex 
quantum computations would require that the laser on the  
$4\,^{2}S_{1/2} - 3\,^{2}P_{5/2}$ 
computation transition have a coherence time as long as the computation time.  
This may necessitate using qubits bridged by Raman transitions as 
discussed above, which 
eliminates the errors caused by the phase drift of the laser. 
\section {Qubit Addressing Optics}
	In order for the $Ca^{+}$ ion qubits to be useful for actual calculations, it 
will be necessary to address the ions in a very controlled 
fashion.  
Our optical system for qubit addressing is shown schematically in 
fig. 6.

\begin{figure}[!ht]
\begin{center}
\epsfxsize=12cm  
\epsfbox{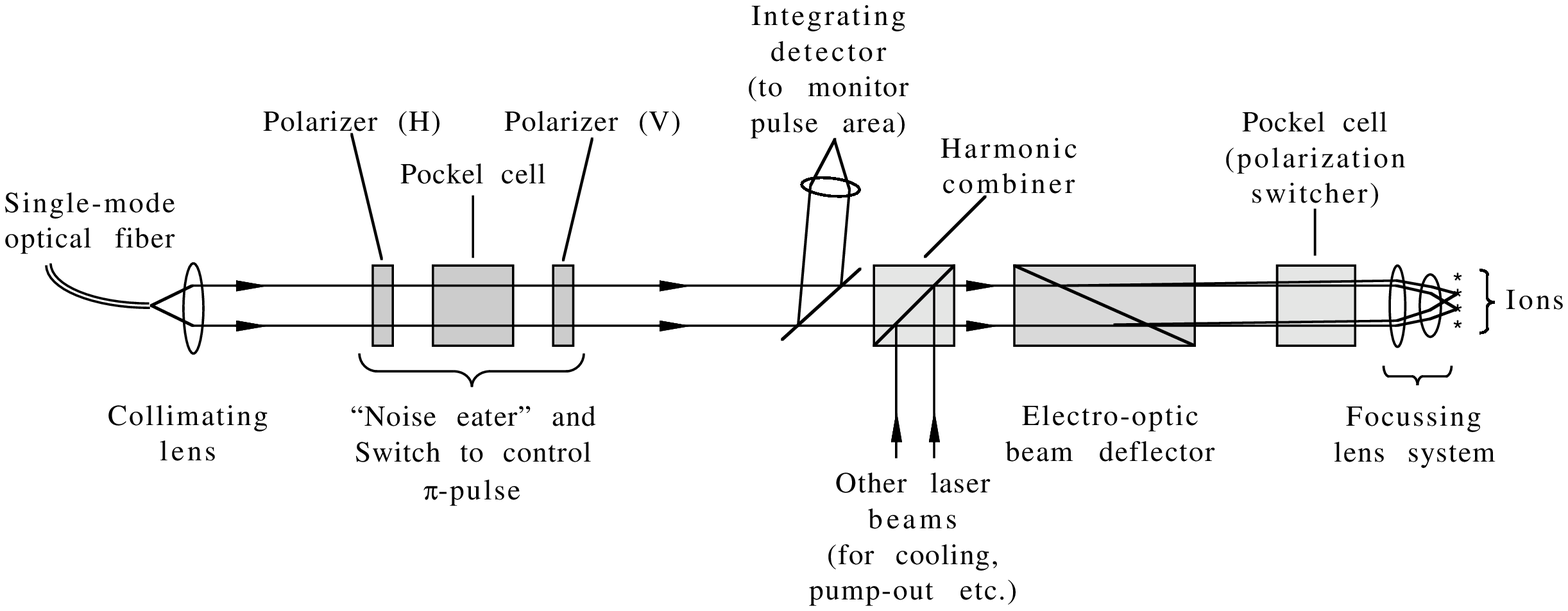}
\end{center}
\caption{Illustration of the laser beam control
optics system.}
\label{figsix}
\end{figure}

There are two aspects to be considered in the
design of such a system: the precise interactions with a 
single ion; and an arrangement for switching between 
different ions in the string.  In addition to the obvious constraints on 
laser frequency and polarization, the primary consideration for making
exact $\pi$- or $2\pi$-pulses is control of the area (over time) of the driving 
light field pulse.  The first step toward this is to stabilize the 
intensity of the laser, as can be done to better than 
$0.1\%$, 
using a standard ``noise-eater".  Such a device typically consists of an 
electro-optical polarization rotator located between two polarizers; the 
output of a fast detector monitoring part of the transmitted beam is used 
in a feedback circuit to adjust the degree of polarization rotation, and 
thus the intensity of the transmitted light.  Switching the light 
beam on and off 
can be performed with a similar (or even the same)
device.  Because such switches can possess rise/fall times on the scale of 
nanoseconds, it should be possible to readily control the area under the 
pulse to within $\sim 0.1\%$, simply by accurately determining the width of the 
pulse.  A more elaborate scheme would involve an integrating detector, 
which would monitor the actual integrated energy under the pulse, shutting 
the pulse off when the desired value is obtained.  

	Once the controls for addressing a single ion are decided, the means for 
switching between ions must be considered.  Any system for achieving this 
must be fast, reproducible, display very precise aiming and low 
``crosstalk" (i.e. overlap of the focal spot onto more than one ion),
and be as simple as possible.  In particular, it is desirable 
to be able to switch between different ions in the string in a time short 
compared to the time required to complete a given $\pi$-pulse on one ion.  
This tends to discount any sort of mechanical scanning system.  
Acousto-optic deflectors, which are often used for similar purposes, may 
be made fast enough, but introduce unwanted frequency shifts on the 
deviated beams.  As a tentative solution, we propose to use an 
electro-optic beam deflector, basically a prism whose index of 
refraction, and consequently whose deflection angle, is varied slightly by 
applying a high voltage across the material; typical switching times for 
these devices is 10 nanoseconds, adequate for our purposes.  
One such device produces a maximum deflection of 
$\pm$9 
mrad, for a $\pm$3000V input.  The associated maximum number of resolvable 
spots (using the Rayleigh criterion) is of order 100, implying that 
$\sim$ 20 
ions could be comfortably resolved with negligible crosstalk.

	After the inter-ion spacing has been determined, i.e., by the trap 
frequencies, the crosstalk specification determines the maximum spot size 
of the addressing beam.  For example, for an ion spacing of 20 $\mu$m, 
any spot size (defined here as the $1/e^2$ diameter) less than 21.6 $\mu$m
will yield a crosstalk of less than 0.1$\%$, assuming a purely Gaussian 
intensity distribution (a good approximation if the light is delivered 
from a single-mode optical fiber, or through an appropriate spatial 
filter).  In practice, scattering and other experimental realities will 
increase this size, so that it is prudent to aim for a somewhat smaller spot 
size, e.g. 10 $\mu$m. One consideration when such small spot sizes are 
required is the effect of lens aberrations, especially since the spot must 
remain small regardless of which ion it is deflected on.  Employing 
standard ray-trace methods, we have found that the blurring effects of 
aberrations can be reduced if a doublet/meniscus lens combination 
is used (assuming an input beam size of 3mm, and an 
effective focal length of 
30mm).	A further complication is that, in order to add or 
remove phonons from the 
system, the addressing beams must have a component along the longitudinal 
axis of the trap. 
The addressing optics must accommodate
a tilted line of focus, otherwise the intensity at each ion would be 
markedly different, and the crosstalk for the outermost ions would become 
unacceptable.  According to ray-trace calculations, adding a simple wedge 
(of $\sim 2^{o}$) solves the problem and this has been
confirmed by measurements using a mock system. 

	Depending on the exact level scheme being considered, it may be necessary 
to vary the polarization of the light.  
Because the electro-optic deflector requires 
a specific linear polarization, any polarization-control elements should 
be placed after the deflector.  The final 
result is a highly directional, tightly-focused beam with controllable 
polarization and intensity.  
\section{Imaging System}
In order to determine the ions' locations and to readout the result
of the quantum computations, an imaging system is required.
Our current imaging system consists of two lenses, one of
which is mounted inside the vacuum chamber, and a video camera coupled to a
dual-stage micro-channel plate (MCP) image intensifier. The first lens with
focal length 15 mm collects the light emitted from the central trap
region with a solid angle of approximately 0.25 sr. The image is relayed
through a 110mm/f2 commercial camera lens to the front plate of the MCP. 
This set-up produces
a magnification of 7.5 at the input of the MCP. The input of the 110 mm lens
is fitted with a 400nm narrow band filter to reduce background from the IR
laser and from light emanating from the hot calcium oven and the electron
gun filament.

The dual plate intensifier is operated at maximum gain for the highest
possible sensitivity. This allows us to read out the camera at normal video
rate of 30 frames ${\rm s}^{-1}$ into a data acquisition computer. Averaging and
integrating of the signal over a given time period can then be undertaken by
software. We find this arrangement extremely useful in enabling us to
observe changes of the cloud size or the intensity of the fluorescence with
changes of external parameters like trapping potential, laser frequency,
laser amplitude, etc. in real time.

The spatial resolution of the system is limited by the active diameter of
individual channels of the MCP of approximately 12 $\mu$m. Since the gain
is run at its maximum value cross talk between adjacent channels in the
transition between the first and second stage is to be expected. This
results in the requirement that two incoming photons can only be resolved
when they are separated at the input of the MCP by at least two channels,
i.e. by 36 $\mu$m in our case. With the magnification of the optical system
of 7.5 this translates into a minimum separation of two ions to be resolved
of 5 $\mu$m, which is well below the separation of ions in the axial well of
about 25 $\mu$m expected in our experiment.
\section{Summary}
It is our contention that currently the ion trap proposal
for realizing a practical quantum computer offers
the best chance of long term success.  This in no way is intended
to trivialize research into the other proposals:
in any of these schemes technological advances 
may at some stage lead to a breakthrough.
In particular, Nuclear Magnetic Resonance does seem to be 
a relatively straightforward
way in which to achieve systems containing a few qubits. 
However, of the technologies which have so far been used
to demonstrate experimental logic gates, ion traps seem to offer the
least number of technological 
problems for scaling up to 10's or even 100's of qubits.

In this paper we have described in some detail the experiment we
are currently developing to investigate the feasibility of
cold trapped ion quantum computation.  We should emphasize
that our intentions are at the moment exploratory: we have
chosen an ion on the basis of current laser technology, rather
than on the basis of which ion which will give the best performance
for the quantum computer. Other species of ion may well give better
performance:  In particular
Beryllium ions do have the
potential for a significantly lower error rate due to spontaneous 
emission, although it is
also true that lighter ions may be more susceptible to heating. 
Other variations, such as the use of Raman transitions in place of
single laser transitions, or the use of standing wave lasers need to 
be investigated. Our choice of Calcium will allow us to 
explore these issues. Furthermore, calculations suggest that it
should be possible to trap 20 or more Calcium ions in a
linear configuration and manipulate their quantum states
by lasers on short enough time scales that many quantum
logic operations may be performed before coherence is lost.
Only by experiment can the theoretical estimates of performance
be confirmed \cite{HJKLP,HJKLPtwo}. Until all of the sources 
of experimental error
in real devices are thoroughly investigated, it will be impossible
to determine what ion and addressing scheme enables one 
to build the best quantum computer 
or, indeed, whether
it is possible to build a useful quantum computer with cold trap ions 
at all.

\section*{Acknowledgments}
This research was funded by the National Security Agency.

\end{document}